\documentclass[ams,preprint]{revtex4}
\usepackage{amsmath}
\usepackage{graphicx}

\newcommand{\be}{\begin{equation}}
\newcommand{\ee}{\end{equation}}
\newcommand{\s}{\cite}
\newcommand{\la}{\label}
\newcommand{\ber}{\begin{eqnarray}}
\newcommand{\eer}{\end{eqnarray}}
\newcommand{\nn}{\nonumber}

\begin{document}
\title{Nonlinear acoustic wave generation in a three-phase seabed}
\newcommand{\mgu}{M.V. Lomonosov Moscow State University, Research Computing Center,\\  Vorobyovy Gory, 
Moscow 119991, Russia}
\author{
A.B. Kukarkin}\email{ jam@srcc.msu.ru}\author{N.I. Pushkina}
\email{ N.Pushkina@mererand.com}\author{ Ya.M. Zhileikin}\email{ jam@srcc.msu.ru}
\affiliation{\mgu} 

\begin{abstract}
     Generation of an acoustic wave by two pump sound waves is studied in a three-phase marine sediment 
that consists of a solid frame and the pore water with air bubbles in it. To avoid  
shock-wave formation the interaction is considered in the frequency range where there is a significant 
amount of sound velocity dispersion. Nonlinear equations 
are obtained to describe the interaction of acoustic waves in the presence of air bubbles. 
An expression for the amplitude 
of the generated wave is obtained and  numerical analysis of its dependence on distance and on 
the resonance frequency of bubbles is performed.  
\end{abstract}
\maketitle

\section{Introduction}

Marine sediments as porous media consisting of a solid frame and the pore water are known to exhibit 
strong elastic nonlinearity  \cite {hov, bel, z1, z2}. 
In the present paper nonlinear sound wave generation by two pump acoustic  waves 
is studied in a three-phase marine sediment, that is in a sediment that contains air bubbles 
in its liquid fraction. 
Nonlinear oscillations of gas bubbles in fluids lead  to considerable enhancement  of the nonlinearity 
of a homogeneous medium \cite{za,wel,zab,wu} as well as of the nonlinear properties of  marine sediments 
that contain bubbles in the pore water \cite{don,ch}.
In Ref. \s {2012} nonlinear interaction of three sound waves was studied in a two-phase marine 
sediment without air bubbles in the liquid phase. In the present paper the sound wave generation 
in a three-phase sediment is considered, 
as in  Ref. \s {2012},  in the  frequency range where there is a 
significant amount of  wave  velocity dispersion \cite {tur,sto,isak,buch}.
Due to velocity dispersion no shock-wave formation would occur which allows observing more explicitly 
interaction of three waves propagating in three different directions. 

\section{Theory}
Our task is to find the amplitude of an acoustic wave $(\omega_3,{\bf k}_3)$ generated by two pump 
acoustic waves 
$(\omega_1, {\bf k}_1)$ and $(\omega_2, {\bf k}_2)$,  the nonlinearity being due to oscillating bubbles. 
The frequencies and wave vectors satisfy the energy and momentum conservation laws,
\[\omega_1+\omega_2=\omega_3,\qquad{\bf k}_1+{\bf k}_2={\bf k}_3.\]

The dependence of the  wave $(\omega_3,{\bf k}_3)$ amplitude on 
the resonance bubble frequency and on distance will be analyzed along with the possibility 
of  experimental observation of the process.

To solve the problem we are based on the well-known Biot model in the form of the equations of 
continuity and momentum conservation for the liquid and solid fractions \s {nik}, 
\ber
\frac{\partial\rho_m}{\partial t}+\rho_{0m}\frac{\partial v}{\partial x}=0,
\,\,\,\,\,\,\,\,\, 
\rho_{0m}\frac{\partial v}{\partial t}=-\frac{\partial p}{\partial x}, \nn\\                                                   
\frac{\partial\rho_s}{\partial t}+\rho_{0s}\frac{\partial u}{\partial x}=0,
\,\,\,\,\,\,\,\,\,
(1-m)\rho_{0s}\frac{\partial u}{\partial t}=         
\frac{\partial \sigma_{xx}}{\partial x}-(1-m)\frac{\partial p}{\partial x}  
\la{1}
\eer

In these equations  we can limit ourselves for simplicity with one-dimensional case since the angles between 
the three waves are small enough. Indeed, 
numerical estimates \s {2012} show that for the experimental 
situation which is used here and which is described in Refs. \cite {tur,buch} the angles between 
the propagation directions of
the waves $(\omega_1, {\bf k}_1),(\omega_2, {\bf k}_2)$ and the  propagation direction
(let it be the $x$-axis) of the wave $(\omega_3,{\bf k}_3)$ are approximately 
$\theta_1\approx25^\circ$ and  $\theta_2 \approx18^\circ.$ 

In  Eqs. (\ref{1}) nonlinear hydrodynamic terms are omitted since as it was noted in Introduction
nonlinear acoustic processes are  governed mainly 
by  nonlinear oscillations of bubbles contained 
in the pore water. In Eqs. (\ref{1}) $\rho_m$  is the mean density of the mixture 
of water and air bubbles;  $\rho_s$ is the solid phase density, 
 (the subscript  "0"  refers to the equilibrium values); $v$ and $u$ are the velocities 
 of the liquid and solid phases; $p$ is the pressure in the water; $m$ is the porosity; 
$\sigma_{xx}$ denotes the effective stress in a porous medium (see \cite{nik}),  
\[\sigma_{xx}=-\left[k+(4/3)\mu\right]\frac{\delta\rho_s}{\rho_{0s}}+
\frac{k}{k_s}p,\]
$k$ and $\mu$ are the bulk and shear moduli of the frame of the porous medium. 

Further calculations performed by one of the authors in Ref. \s{eng} lead to the equations   
\ber
\frac{\partial^2p}{\partial t^2}-\frac{m}{\rho_{0f}G}\frac{\partial^2p}
{\partial x^2}-\frac{\nu}{\rho_{0s}G}\frac{\partial^2\rho_s}{\partial
 t^2}=\frac{mn}{G}\frac{\partial^2V}{\partial t^2}\nn\\
\frac{\partial^2\rho_s}{\partial t^2}(1-m)-
\frac{k+(4/3)\mu}{\rho_{0s}}\frac{\partial^2\rho_s}{\partial x^2}
 -\nu\frac{\partial^2p}{\partial x^2}=0 \la{2}
 \eer
Here $n$ is the bubble concentration and $V$ is the bubble volume.
Eqs. (\ref{2}) should be supplemented with the equation for the individual bubble motion \s{zab},
\be
\ddot V+\omega_0^2V+f\dot V-\alpha V^2-\beta(2\ddot VV+\dot V^2)=
\epsilon p, \la{3}
\ee  
where $\omega_0$ is the resonance bubble frequency. The coefficients in Eq. (\ref{3})
are expressed through the equilibrium bubble volume $V_0$, its radius $R_0$ and 
the adiabatic index  $\gamma$,
\[\alpha =\omega_0^2(1+\gamma)/2V_0,\,\,\,\,\,\,\,\beta =1/6 V_0,\,\,\,\,\,\,
\epsilon =4\pi R_0/\rho_{0f},\,\,\,\,\,\,\,f=\delta\omega_0,\]
 $\delta$ is
 the dimensionless absorption coefficient of  bubble oscillations.

As  was mentioned in Introduction in order to observe three-wave interaction in a more explicit way we 
should use the frequency range with a noticeable velocity dispersion. It is preferable to choose
the interval of the maximum velocity dispersion to ensure that the angles between the waves 
are not too small, otherwise they would fall in the dissipation spreading of the waves. 
As in Ref. \s{2012}, the data from 
\s {tur} listed in Fig. 2 of \s {buch} will be used. Basing on these data we choose the frequencies 
equal to $\omega_1=2\pi\cdot2\cdot10^3\,{\rm s}^{-1}$,  $\omega_2=2\pi\cdot3\cdot10^3\,{\rm s}^{-1}$, 
for which 
the sum frequency equals\,\, 
$2\pi\cdot5\cdot10^3\,{\rm s}^{-1}$. 
We shall seek for the solution to Eqs. (\ref{2}), (\ref{3}) in the form
\be
p_i=1/2[P_ie^{i(k_ix-\omega_it)}+{\rm c.c.}], \,\,\,\,\,\,\,\,\,\,V_i=1/2[V_ie^{i(k_ix-\omega_it)}+
{\rm c.c.}],\,\,\,\,\,\,i=1,2,3. \nn 
\ee 
 
 The only nonlinear term in Eqs. (\ref{2}) that ensures  the nonlinear process 
is the last term of the first equation.
In fact this term is the sum of the linear $V^l$ and the nonlinear part  $V^n$.
 The nonlinear part $V^n$ responsible for the generation of the wave $p_3$ 
by the waves $p_1$, $p_2$  can be obtained from Eq. (\ref{3}) in the form 
\be
V^n=\frac{[\alpha-\beta(\omega_1^2+\omega_2^2+\omega_1\omega_2)]\epsilon^2p_1p_2}
{(-\omega_3^2+\omega_0^2-i\omega_3f)
(\omega_1^2-\omega_0^2+i\omega_1f)(\omega_2^2-\omega_0^2+i\omega_2f)}. \la{4}  
\ee

In solving the equations (\ref{2}) the so called approximation of a fixed field is used, that is 
the changes of the amplitudes $P_1$, $P_2$ because of nonlinearity are neglected, their changes being only 
due to linear bubble oscillations that cause dispersion in the medium.   
Using (\ref{4}) we obtain the solution to Eqs. (\ref{2}) for the amplitudes $P_1, P_2, P_3$ that are assumed 
to vary slowly at the distance of order of the wave length, 
\ber
\frac{dP_1}{dx}=ib_1P_1,\,\,\,\,\,\,\,\,\,
\frac{dP_2}{dx}=ib_2P_2,\,\,\,\,\,\,\,\,\, 
\frac{dP_3}{dx}=ib_3P_3-iaP_1P_2. \la{5}
\eer
The following notations are used in Eqs. (\ref{5}),
\[b_j=(\rho_fnV\omega_jcK/2\gamma P_0L)\left(1-\omega_j^2/\omega_0^2-i\delta_j\omega_j^2/
\omega_0^2\right)^{-1},\,\,\,\,\,\,j=1,2,3,\]
where
\[K=1-m-\frac{k+(4/3)\mu}{\rho_sc^2},\,\,\,\,\,\,\,\,\,
L=K-\frac{\nu^2}{m}\frac{\rho_f}{\rho_s}\left[\frac{k+(4/3)\mu}{\rho_sc^2}K^{-1}+1\right],\,\,\,\,\,\,\,\,\,\, 
\nu=1-m-k/k_s.\]
$P_0$ is the equilibrium pressure in the pore water, related to other equilibrium parameters 
of a bubble as $\omega_0^2=3\gamma P_0/\rho_0R_0^2$.  
The quantity $a$ in the last of Eqs. (\ref{5}) is the vertex of the nonlinear interaction, it equals 
\[a=\frac{\rho_fnV\omega_3cK\left[1+\gamma-(1/3\omega^2_0)
(\omega_1^2+\omega_2^2+\omega_1\omega_2)
\right]}{4\gamma^2P_0^2L(1-\omega_1^2/\omega_0^2-i\delta_1\omega_1^2/\omega_0^2)(1-\omega_2^2/\omega_0^2-
i\delta_2\omega_2^2/\omega_0^2)(1-\omega_3^2/\omega_0^2-i\delta_3\omega_3^2/\omega_0^2)}\]

As can be seen from Eqs. (\ref{5})  the imaginary 
parts of the quantities $b_j$ define  dissipation of the acoustic waves. This dissipation is caused by
energy losses due to bubble oscillations. It  grows significantly when the resonance bubble 
frequency  $\omega_0$ approaches one of the frequencies of the interacting waves. Numerical estimates show 
that in the frequency range 
where $\omega_0$ is either much higher or much lower than the frequencies of the acoustic 
waves the wave attenuation caused by bubble oscillations is considerably less than   
acoustic dissipation of hydrodynamic nature. This dissipation  does not enter Eqs. (\ref{5}), but it is to 
be taken into 
account to estimate the real  distance of the nonlinear interaction. The distance 
 cannot exceed the dissipation lengths of the 
waves $(\omega_1, {\bf k}_1)$ and $(\omega_2, {\bf k}_2)$. The amplitude attenuation coefficients 
of these waves are equal correspondingly to 
$\alpha_1\approx 0.8\cdot10^{-3}\,{\rm cm}^{-1}$ 
and to $\alpha_2\approx3\cdot10^{-3}\,{\rm cm}^{-1}$ (see Ref. \cite{buch}). 
This corresponds approximately to 
propagation distances $\sim1250 \,{\rm cm}$ and $\sim330 \, {\rm cm}$ which means, 
that the interaction length does not  exceed the distance that is the order of  
$300 \, {\rm cm}$. 

If at $x=0$ the amplitude $|P_3|$ 
of the generated wave is at the fluctuation level, that is practically zero, the solution to Eqs. 
(\ref{5}) can be represented in the form  
\begin{equation}
|P_3|^2=\omega_3^2|P_1|^2|P_2|^2E\left[1+\gamma-(1/3\omega^2_0)(\omega_1^2+\omega_2^2+\omega_1\omega_2)
\right]^2/4\gamma^2P_0^2D,  \label{6}
\end{equation}
with 
\[E(x)=e^{-2{\rm Im}b_3x}+e^{-2{\rm Im}(b_1+b_2)x}-2e^{-2{\rm Im}(b_1+b_2+b_3)x}
\cos[{\rm Re}(b_1+b_2-b_3)x],\]
\begin{equation}
D=\omega_1^2\Theta_2\Theta_3+\omega_2^2\Theta_1\Theta_3+\omega_3^2\Theta_1\Theta_2+
2\omega_1\omega_2\Theta_3\Phi_{12}-2\omega_1\omega_3\Theta_2\Phi_{13}-2\omega_2\omega_3\Theta_1\Phi_{23}.
\label{7}
\end{equation}
In the formula (\ref{7}) the following notations are used,
\[\Theta_j=
\alpha^2_j+\beta_j^2,\,\,\,\,\,\,\,\,\,\,\,\,\Phi_{jk}=\alpha_j\alpha_k+\beta_j\beta_k,\,\,\,\,\,\,\,\,
j,k=1,2,3,\]
where 
\[\alpha_{j}=1-\omega_{j}^2/\omega_0^2,\,\,\,\,\,\,\,\,\,\,\,
\beta_{j}=\delta_{j}\omega_{j}^2/\omega_0^2.\]

We shall analyze the obtained amplitude of the generated wave, namely its dependence  
 on distance at different resonance bubble frequencies and on the resonance bubble frequency at a 
given distance.
To perform this analysis and for numerical estimates the following quantities and experimental data
will be used \cite{med,buch,tur},
$$
P_0\approx3\times10^6\,{\rm dyn/cm}^2,\,\,\,\,\,\,\,\,\,P_1=P_2\approx10^5\,{\rm dyn/cm}^2,\,\,\,\,\,\,\,\,
\gamma=1.4,
\,\,\,\,\,\,\,\,\ nV=10^{-5},$$
$$\omega_1=4\pi\times10^3\,{\rm s}^{-1},\,\,\,\,\,\,\,\omega_2=6\pi\times10^3\,{\rm s}^{-1};\,\,\,\,\,\,\,
\omega_3=10\pi\times10^3\,{\rm s}^{-1},\,\,\,\,\,\,\,\,\delta_j\approx4\times10^{-2},$$
$$\rho_f=1\,{\rm g/cm}^3,\,\,\,\,\,\,\,\,\,\rho_s=2.65\,{\rm g/cm}^3, \,\,\,\,\,\,\,\,\,
m=0.4, \,\,\,\,\,\,\,\,
c\approx1.7\times10^5\,{\rm cm/s}, $$
$$k=10^9{\rm dyn/cm}^2,\,\,\,\,\,\,\,\,\,\,\mu=5\times10^8\,{\rm dyn/cm}^2,
\,\,\,\,\,\,\,\,\, k_s=3.6\times10^{11}\,{\rm dyn/cm}^2.$$

Since the obtained formula (\ref{6}) is somewhat complicated computer simulation will be applied 
to investigate the behaviour of the  sound-wave amplitude $|P_3|$ for various 
resonance bubble frequencies and various distances. 

1. The dependence of the ratio $|P_3|/|P_1|$ on distance  
   for the resonance bubble frequency  $\omega_0$ that is less than the  frequencies of the three waves, 
   let it be $\omega_0=4\pi\times10^2\,{\rm s}^{-1}$. The distance changes in the interval 
   $100\,{\rm cm}<x<300\,{\rm cm}$.
   In this case  the value of $|P_3|$ remains at any distance several orders of magnitude less than $|P_1|$.
     
2. The resonance bubble frequency  equals one of the frequencies of the three waves. The dependence 
   of the ratio $|P_3|/|P_1|$ on 
   distance in the interval $100\,{\rm cm}<x<300\,{\rm cm}$ is the following, 

   a) $\omega_0=\omega_1$. The ratio  $|P_3|/|P_1|$ is always rather small and is equal to $\approx(3.6-3.5)
   \times10^{-3}$. 
   
   b) $\omega_0=\omega_2$. The ratio  $|P_3|/|P_1|$ becomes higher, of the order of $\approx3\times
   10^{-2}$ and declines a little with distance because of dissipation due to bubble oscillations.
   
   c) $\omega_0=\omega_3$. The ratio  $|P_3|/|P_1|$ is a little higher than in the two preceding 
   cases, it is  of the order of
   $\approx4\times10^{-2}$, slightly decreasing with distance. 
   
   The decrease of the ratio $|P_3|/|P_1|$ with distance in the last three cases is associated with 
   the fact that the 
   resonance bubble frequency in these cases always falls in resonance with one of the interacting 
   waves which leads 
   to enhancement of the energy losses due to bubble oscillations. On the other hand this evidences 
   as well that in this frequency range nonlinear interaction is too weak to overcome dissipation.
   
3. The resonance bubble frequency  $\omega_0$ is higher than the frequencies of the three waves, let it be 
    equal  to $10\pi\times10^4\,{\rm s}^{-1}$. 
   The distance changes in the interval $100\,{\rm cm}<x<300\,{\rm cm}$. In this case the amplitude $|P_3|$ 
   grows slowly starting from the  value $\approx0,17 |P_1|$ up to the value $\approx0.3 |P_1|$. 
   This means that nonlinear oscillations of bubbles with  high resonance frequencies 
   ensures an effective generation of the sum frequency sound wave.
   
In Fig. 1 the dependence of $|P_3|/|P_1|$ on the ratio $\omega_0/\omega_3$ at the distance $x=300\,\,{\rm cm}$ 
is presented for the whole range of 
 $\omega_0$ starting from the frequency  much lower than the frequencies of the interacting acoustic 
 waves up to the frequencies that are much higher than the frequencies $\omega_1$, $\omega_2$, and 
 $\omega_3$. 
For $\omega_0$ just between the frequencies  $\omega_1$ and $\omega_2$ there is a narrow peak that  
 can be attributed to the fact that at this point $\omega_0$ is equally 
far from the resonances with  $\omega_1$ and $\omega_2$. For  $\omega_0>\omega_3$ there starts 
practically steady rise of the amplitude of the generated wave $|P_3|$ up to the value     
$|P_3|\approx0.3 |P_1|$.

\bigskip

\begin{figure}
\begin{center}
\includegraphics[scale=0.4]{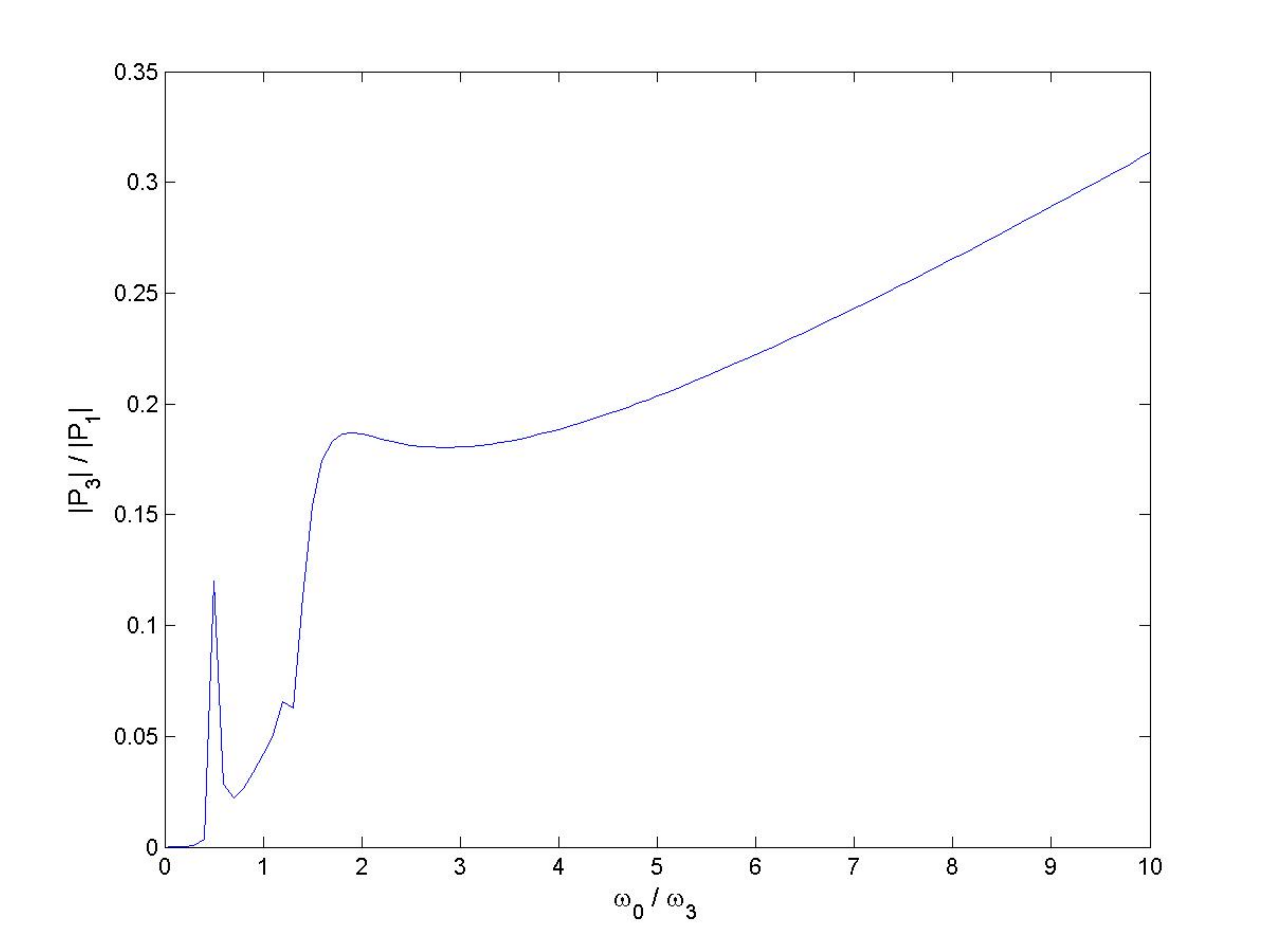}
\caption{The dependence of $|P_3|/|P_1|$ on $\omega_0/\omega_3$ at the distance $x=300\,{\rm cm}$ 
(the interaction length).}
\end{center}
\end{figure}

\section{Summary}

The presented results show that the considered nonlinear acoustic process  "feel" very slightly 
bubble oscillations  with low
resonance frequencies. Instead bubble oscillations with high resonance frequencies influence significantly 
nonlinear  wave interaction. 
The generated wave amplitude in this case  achieves a measurable value of the tenths of the pump amplitudes 
at a reasonable distance.  
The range where  $\omega_0$ coincides with one of the frequencies of the
interacting waves is not favorable for nonlinear acoustic interaction not only because 
of higher energy losses when bubble oscillations fall in resonance with the interacting acoustic waves, 
but also because $\omega_0$ is not sufficiently high in that interval.

\newpage

\end{document}